\documentclass[12pt]{article}
\usepackage{graphicx}
\usepackage{a4p}
\usepackage{amssymb}
\usepackage{amsmath}
\def\sqee{\ifmmode{\sqrt{s_\text{ee}}}\else
  {$\sqrt{s_\text{ee}}$}\fi}
\def\sqeeb{\ifmmode{\sqrt{s_{\protect\bf\text{ee}}}}\else
  {$\sqrt{s_{\protect\bf\text{ee}}}$}\fi}
\def\kp{{\ifmmode{k_{\perp}}\else{$k_{\perp}$}\fi}}
\def\ptjet{\ifmmode{p^\text{jet}_\text{T}}\else 
  {$p^\text{jet}_\text{T}$}\fi}
\def\pt{\ifmmode{p_\text{T}}\else 
  {$p_\text{T}$}\fi}
\def\etajet{\ifmmode{|\eta^\text{jet}|}\else 
  {$|\eta^\text{jet}|$}\fi}
\def\etaj{\ifmmode{\eta^\text{jet}}\else 
  {$\eta^\text{jet}$}\fi}
\def\mj1h2{\ifmmode{M_{\text{J1H2}}}\else 
  {$M_{\text{J1H2}}$}\fi}
\def\gsg{\ifmmode{\gamma^{\star}\gamma}\else
  {$\gamma^{\star}\gamma$}\fi}
\def\epem{\ifmmode{\text{e}^{+}\text{e}^{-}}\else
  {$\text{e}^{+}\text{e}^{-}$}\fi}
\def\as{\ifmmode{\alpha_{s}}\else
  {$\alpha_{s}$}\fi}
\def\avptjet{\ifmmode{\langle \ptjet \rangle}\else
  {$\langle \ptjet \rangle$}\fi}
\begin{document}
\begin{titlepage}

\begin{center}
{\large EUROPEAN ORGANISATION FOR NUCLEAR RESEARCH}
\end{center}
\vspace*{1.0cm}

\begin{flushright}
CERN-PH-EP-2007-016\\
3rd May 2007
\end{flushright}
\vspace*{1.0cm}

\begin{center}
{\boldmath\LARGE\bf Inclusive Jet Production in Photon-Photon \\
\vspace*{0.3cm}
Collisions at $\sqeeb$ from $189$ to $209$~GeV \unboldmath}
\end{center}
\vspace*{0.5cm}

\begin{center} {\LARGE The OPAL Collaboration} \end{center}
\vspace*{0.5cm}

\begin{center}
{\large  Abstract}
\end{center}

Inclusive jet production ($\epem \rightarrow \epem \text{+jet+X}$) is
studied in collisions of quasi-real photons radiated by the LEP beams
at {\epem} centre-of-mass energies {\sqee} from 189 to 209~GeV.  Jets
are reconstructed using the {\kp} jet algorithm. The inclusive
differential cross-section is measured as a function of the jet
transverse momentum, {\ptjet}, in the range 5~$< \ptjet <$~40~GeV for
pseudo-rapidities, {\etaj}, in the range -1.5~$< \etaj <$~1.5. The
results are compared to predictions of perturbative QCD in
next-to-leading order in the strong coupling constant.

\vfill
\begin{center}
{\large (Submitted to Physics Letters B)}
\end{center}
\end{titlepage}
\begin{center}{\Large        The OPAL Collaboration
}\end{center}\bigskip
\begin{center}{
G.\thinspace Abbiendi$^{  2}$,
C.\thinspace Ainsley$^{  5}$,
P.F.\thinspace {\AA}kesson$^{  7}$,
G.\thinspace Alexander$^{ 21}$,
G.\thinspace Anagnostou$^{  1}$,
K.J.\thinspace Anderson$^{  8}$,
S.\thinspace Asai$^{ 22}$,
D.\thinspace Axen$^{ 26}$,
I.\thinspace Bailey$^{ 25}$,
E.\thinspace Barberio$^{  7,   p}$
% ...
}\end{center}\bigskip
\bigskip
$^{  1}$School of Physics and Astronomy, University of Birmingham,
Birmingham B15 2TT, UK
\newline
$^{  2}$Dipartimento di Fisica dell' Universit\`a di Bologna and INFN,
I-40126 Bologna, Italy
\newline
% ...
$^{  5}$Cavendish Laboratory, Cambridge CB3 0HE, UK
\newline
% ...
$^{  7}$CERN, European Organisation for Nuclear Research,
CH-1211 Geneva 23, Switzerland
\newline
$^{  8}$Enrico Fermi Institute and Department of Physics,
University of Chicago, Chicago IL 60637, USA
\newline
% ...
$^{ 21}$Department of Physics and Astronomy, Tel Aviv University,
Tel Aviv 69978, Israel
\newline
$^{ 22}$International Centre for Elementary Particle Physics and
Department of Physics, University of Tokyo, Tokyo 113-0033, and
Kobe University, Kobe 657-8501, Japan
\newline
% ...
$^{ 25}$University of Victoria, Department of Physics, P O Box 3055,
Victoria BC V8W 3P6, Canada
\newline
$^{ 26}$University of British Columbia, Department of Physics,
Vancouver BC V6T 1Z1, Canada
\newline
% ...
$^{  p}$ now at The University of Melbourne, Victoria, Australia
\newline
% ...
\newpage
\section{Introduction}
\label{sec:intro}

We have studied the inclusive production of jets in collisions of two
quasi-real photons at {\epem} centre-of-mass energies {\sqee} from 189
to 209~GeV, with a total integrated luminosity of 593~$\text{pb}^{-1}$
collected by the OPAL detector at LEP. The transverse
momentum\footnote{OPAL uses a right-handed coordinate system where the
$z$-axis points in the direction of the e$^-$ beam and the $x$-axis
points to the centre of the LEP ring. The polar angle $\theta$ and the
azimuthal angle $\phi$ are defined relative to the $+z$-axis and
$+x$-axis, respectively.  In cylindrical polar coordinates, the radial
coordinate is denoted $r$. The transverse momentum is defined as the
component of the momentum perpendicular to the $z$-axis. The
pseudo-rapidity {$\eta$} is defined as $\eta=-\ln\tan(\theta/2)$.} of
the jets provides a scale that allows such processes to be calculated
in perturbative QCD. Calculations at next-to-leading order (NLO) in
the strong coupling constant, {\as}, for this process are available
\cite{bib:klasen, bib:frixione}. Comparisons of these calculations to
the data presented in this paper provide tests of perturbative QCD for
the observables and kinematical region considered.  Leading order (LO)
Monte Carlo (MC) generators are used to estimate the influence of soft
underlying processes not included in the NLO calculation. Furthermore
the measured jet cross-sections may be used in studies evaluating the
hadronic structure of the photon, which are beyond the scope of this
paper.

The jets are reconstructed using the {\kp} jet
algorithm~\cite{bib:ktclus}.  Inclusive jet cross-sections in
two-photon interactions have previously been measured at {\sqee} from
130 to 136~GeV by OPAL \cite{bib:opalonejetpaper}, and at {\sqee} from
189 to 209~GeV by L3 \cite{bib:l3paper}.  In \cite{bib:l3paper} an
excess at high transverse momenta was observed in the data over the
QCD calculations, for kinematical conditions very similar to those
used in the present paper.

At LEP the photons are radiated by the beam
electrons\footnote{Positrons are also referred to as electrons.}, and
carry mostly small negative four-momenta squared, $Q^2$.  In this
paper events are considered only if the electrons are scattered at
small angles and are not detected. Both photons are therefore
quasi-real ($Q^2 \approx 0$~GeV$^2$). The interactions can be modelled
by assuming that each photon can either interact directly or appear
resolved through its fluctuations into hadronic states.  In leading
order QCD this model leads to three different event classes for
$\gamma\gamma$ interactions: direct, single-resolved and
double-resolved, where resolved means that the incoming photon
interacts through its partonic structure (quarks or gluons).


\section{The OPAL detector}

A detailed description of the OPAL detector can be found
elsewhere~\cite{bib:opaltechnicalpaper}.  The central tracking was
performed inside a solenoidal magnet which provided a uniform axial
magnetic field of 0.435~T along the beam axis.  Starting from the
innermost components, the tracking system consisted of a high
precision silicon micro-vertex detector, a precision vertex drift
chamber, a large volume jet chamber with 159 layers of axial anode
wires and a set of $z$ chambers measuring the track coordinates along
the beam direction.  The transverse momenta, $\pt$, of tracks were
measured with a precision parametrised by
$\sigma_{\pt}/\pt=\sqrt{0.02^2+(0.0015\cdot \pt)^2}$ ($\pt$ in GeV) in
the central region $|\cos\theta|<0.73$. The position of the primary
vertex is determined from the tracks.

The magnet was surrounded in the barrel region ($|\cos\theta|<0.82$)
by a lead-glass electromagnetic calorimeter (ECAL) and a hadronic
sampling calorimeter (HCAL), which in turn were surrounded by muon
chambers. Similar layers of detectors were installed in the endcaps
($0.82<|\cos\theta|<0.98$).  The small-angle region from 47 to
140~mrad around the beam pipe on both sides of the interaction point
was covered by the forward calorimeters (FD) and the region from 33 to
59~mrad by the silicon-tungsten luminometers (SW). The latter were used
to determine the luminosity by counting small-angle Bhabha scattering
events.

\section{Kinematics and MC simulation}
\label{sec:datamc}

The properties of the two interacting photons ($i=1,2$) are described
by their negative squared four-momenta $Q_{i}^2$ and the invariant
mass of the photon-photon system, $W$. Each $Q_i^2$ is related to the
electron scattering angle $\theta'_i$ relative to the beam axis
by
\begin{equation}
Q_i^2 = -(p_i-p'_i)^2\approx 2E_i E'_i(1-\cos\theta'_i),
\label{eq-q2}
\end{equation}
where $p_i$ and $p'_i$ are the four-momenta of the beam electrons and
the scattered electrons, respectively, and $E_i$ and $E'_i$ are their
energies.  Events with one or both scattered electrons detected
(single-tagged or double-tagged events) are excluded from the
analysis. Driven by the angular acceptance of the FD and SW
calorimeters, a value of $Q^2=4.5$~GeV$^2$ is used in this analysis as
the maximum possible $Q^2$. The median $Q^2$ resulting from this limit
cannot be determined from data, since the scattered electrons are not
tagged, but is predicted by MC simulations to be of the order of
$10^{-4}$~GeV$^2$. The invariant mass of the photon-photon system, $W$, can be
obtained from the energies and momenta $(E_{\rm h} , \vec{p}_{\rm h})$
of the final state hadrons.

All signal and background Monte Carlo samples used for detector
corrections and background determinations were passed through a
full simulation of the OPAL detector \cite{bib:gopal}. They are
analysed using the same reconstruction algorithms as are applied to
the data.
 
The Monte Carlo generator PYTHIA~5.722 \cite{bib:pythia,bib:schuler}
was used to simulate the signal processes for the determination of
detector corrections, as large samples with full detector simulation
and reconstruction were available. For all other purposes the more
modern PYTHIA~6.221 was used to generate signal samples.
PYTHIA is based on LO~QCD matrix elements for massless
quarks with the addition of parton showers and hadronisation.
The following generators were used for the simulation of the six
background processes that contribute significantly after the event
selection described below: PYTHIA for $\rm{Z}/\gamma^\star \rightarrow
\text{q}\bar{\text{q}}$ and $\text{e}^+\text{e}^- \rightarrow
\text{W}^+\text{W}^-$; BDK \cite{bib:bdk} for $\gamma\gamma
\rightarrow \tau^+\tau^-$; HERWIG \cite{bib:herwig} for deep-inelastic
electron-photon scattering ({\gsg}); KORALZ \cite{bib:koralz} for $\rm{Z}
\rightarrow \tau^+\tau^-$ and GRC4F \cite{bib:grc4f} for
$\text{e}^+\text{e}^- \rightarrow
\text{e}^+\text{e}^-\text{q}\bar{\text{q}}$ events.

\section{Jet definition and event selection}
\label{sec:eselect}
The data presented were collected by the OPAL detector at
centre-of-mass energies $\sqee =$ 189~-~209~GeV and represent a total
integrated luminosity of 593~$\text{pb}^{-1}$.  For the purpose of
this analysis, the difference between the data taken at the various
values of {\sqee} is small and therefore the distributions for all
energies have been added. The luminosity-weighted average
centre-of-mass energy is 198.5~GeV.  The efficiency to
trigger jet events in the selected kinematical region is close to
100\% \cite{bib:opalonejetpaper}.

In this analysis, a sum over all particles in the event or in a jet
means a sum over two kinds of objects: tracks satisfying the quality
cuts detailed below, and all calorimeter clusters, including those in
the FD and SW calorimeters. A track is required to have a minimum
transverse momentum of 120~MeV and at least 20 hits in the central jet
chamber.  The point of closest approach to the origin must have a
distance of less than 25~cm in $z$ and a radial distance, $d_0$, of
less than 2~cm from the $z$-axis.  For tracks with a transverse
momentum larger than 5~GeV, $d_0$ is required to be less than 0.15~cm,
to ensure a good momentum measurement.  Calorimeter clusters have to
pass an energy threshold of 100~MeV in the barrel section or 250~MeV
in the endcap section of the ECAL, 600~MeV for the barrel and endcap
sections of the HCAL, 1~GeV for the FD, and 2~GeV for the SW.  An
algorithm is applied to avoid double-counting of particle momenta in
the central tracking system and their energy deposits in the
calorimeters~\cite{bib:opalonejetpaper}. The measured hadronic final
state for each event consists of all objects thus defined.

Events with at least one jet are first preselected before the final
event selection based on maximum likelihood distribution
functions~\cite{bib:pc} is applied. The preselection criteria are as
follows.

\begin{itemize}

\item Using the {\kp} jet algorithm, the
  event must contain at least one jet with $\etajet<1.5$ and a
  transverse momentum $\ptjet>5$~GeV.  In this algorithm the distance
  measure between any pair of objects $\{ i, j \}$ to be
  clustered is taken to be
  \begin{equation} 
     d_{ij} = \min(p^2_{Ti}, p^2_{Tj}) (R^2_{ij}/R^2_0)
     ~~~\mathrm{with}~~~ R^2_{ij} = (\Delta\eta_{ij})^2 +
     (\Delta\phi_{ij})^2.
  \end{equation}
  Throughout this analysis we set $R^2_0 = 1$. The $p_T$ of the
  reconstructed jet, {\ptjet}, is calculated as the sum of the transverse
  momenta of all the particles in the jet.

\item The total summed energy deposited in the ECAL and the HCAL has
  to be less than 80~GeV. This removes most of the hadronic $\rm{Z}$
  decays, including events with a radiative return to the $\rm{Z}$
  peak.

\item To remove events with scattered electrons in the FD or in the SW
  calorimeters, the total energy sum measured in the FD has to be less
  than 55~GeV, and the total energy sum measured in the SW calorimeter
  has to be less than 40~GeV.
      
\item The $z$ position of the primary vertex is required to satisfy
  $|z|<5$~cm and the net charge $Q$ of the event calculated from
  adding the charges of all tracks is required to be $|Q|<5$ to reduce
  background due to beam-gas interactions.

\item To remove events originating from interactions between beam
  electrons and the beam-pipe the radial distance of the primary
  vertex from the beam axis has to be less than 3~cm.

\item The event must lie in the allowed ranges for the input variables
  of the maximum likelihood selection, as detailed below.

\end{itemize}

The final event selection uses seven input variables for the likelihood function:

\begin{enumerate}

\item The visible invariant mass measured in the ECAL only,
  $W_{\text{ECAL}}$ (in the range \mbox{[0-80]~GeV});

\item The visible invariant mass calculated from the entire hadronic
  final state, $W_{\text{rec}}$ (in the range \mbox{[0-120]~GeV});

\item The number of tracks (in the range \mbox{[6-70]});

\item The sum of all energy deposits in the ECAL (in the range
  \mbox{[0-80]~GeV});

\item The sum of all energy deposits in the HCAL (in the range
  \mbox{[0.1-55]~GeV});

\item The missing transverse momentum of the event
  calculated from the measured hadronic final
  state (in the range from zero to $\sqee/2$);

\item To improve the rejection of background coming from hadronic
  $\rm{Z}$ decays, an
  invariant mass, \mj1h2, is calculated from the jet with highest
  $\ptjet$ in the event and the four-vector sum of all hadronic final
  state objects in the hemisphere opposite to the direction defined by
  this jet (considered in the range
  \mbox{[0.1-100]~GeV}).

\end{enumerate}

In comparing the preselected events to MC simulations, 
the signal MC generator PYTHIA~5.722 underestimates the normalisation of the
cross-section by about 50\% in this process, and is scaled up
accordingly. A similar deficiency was also observed in our previous
study on di-jets~\cite{bib:dijets}. Furthermore previous studies have
shown that the prediction of MC generators for jet events in
photon-photon collisions where one of the photons is virtual is too
low by about a factor of two \cite{bib:rooke}. The prediction of the
contribution from {\gsg} events has been scaled up accordingly,
resulting in an adequate description of all quantities used in the
analysis.

\begin{figure}[ht]
\begin{center}
\includegraphics[width=0.95\textwidth]{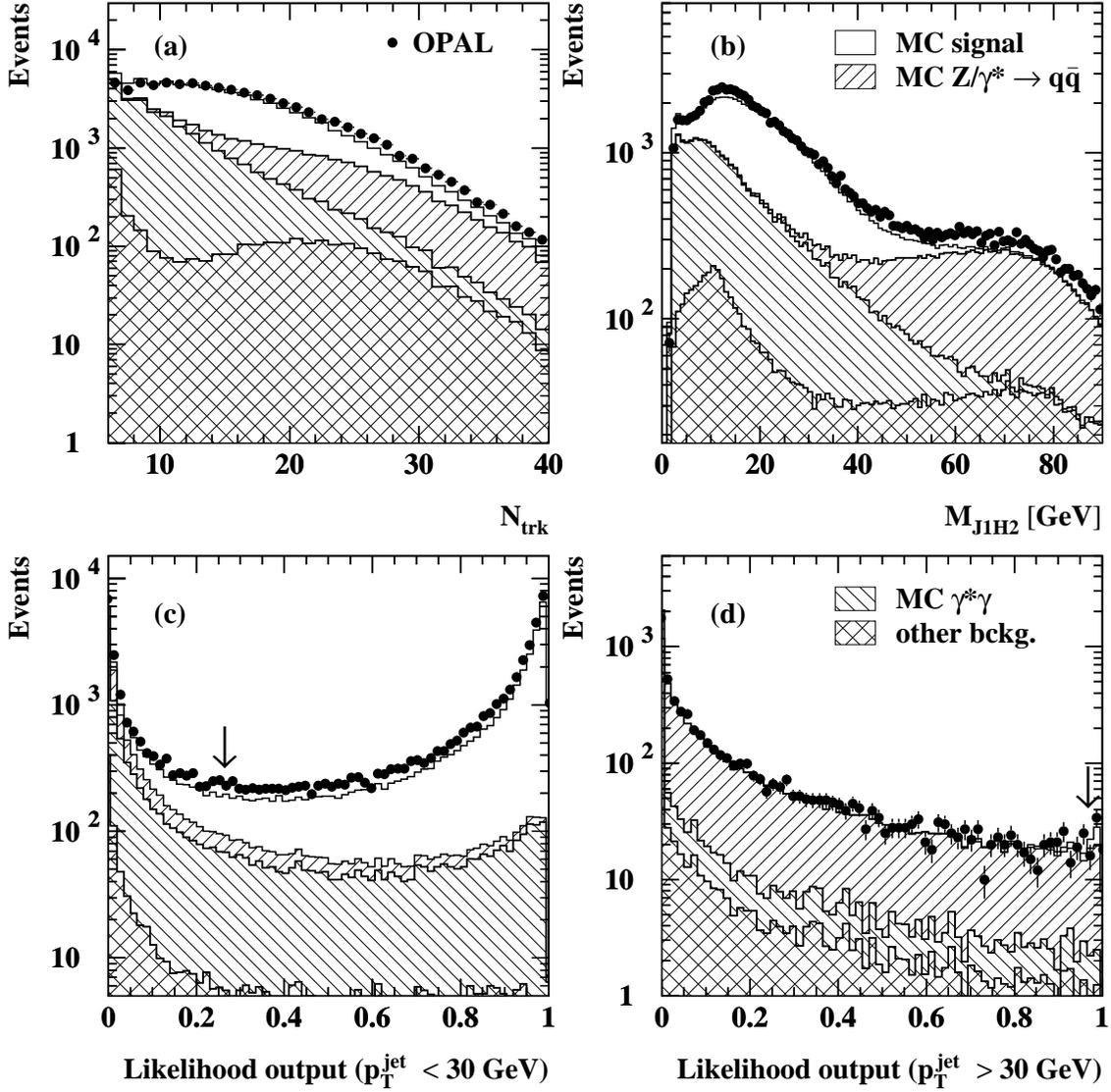}
\end{center}
\caption{ Example inputs to the likelihood functions: (a) shows the
   number of tracks in the event, and (b) the invariant mass of the
   system formed by the jet with the highest $\ptjet$ in the event and
   the four-vector calculated from all objects in the opposite
   hemisphere as seen from this jet. Outputs of the likelihood
   functions: Plots (c) and (d) show the output of the likelihood
   functions for events with $\ptjet <$~30~GeV and $\ptjet >$~30~GeV,
   respectively. Events are selected with likelihood values larger than
   the cuts indicated by the arrows. The signal MC and the
   contribution of the $\gamma^\star\gamma$ MC have been scaled up as
   described in section~\ref{sec:eselect}.}
\label{fig:likinoutput}
\end{figure}

Figure~\ref{fig:likinoutput} (a) and (b) show two examples of the
input distributions used for the likelihood selection, which is
performed separately for events with the highest {\ptjet} smaller than
30\,GeV and events with the highest {\ptjet} larger than 30\,GeV. The
region of high {\ptjet}, where most of the discrepancy with NLO QCD is
observed in~\cite{bib:l3paper}, is strongly affected by background
from $\rm{Z}/\gamma^\star \rightarrow \text{q}\bar{\text{q}}$ which is
not important at lower {\ptjet}. A separate optimization of the
selection is hence necessary to maximize the reach of the analysis in
{\ptjet}.  The output of the likelihood functions for the data and all
simulated processes is shown in Figure~\ref{fig:likinoutput} (c) and
(d).  The cuts on the likelihood outputs are chosen to be 0.26 and
0.98 for the low and high {\ptjet} region, respectively.  Applying
these cuts reduces the background by 99.5\% while reducing the signal
by 71\% in the high {\ptjet} region; in the low {\ptjet} region, these
reductions are 91\% and 27\%, respectively.

\begin{figure}[ht]
\begin{center}
\includegraphics[width=0.95\textwidth]{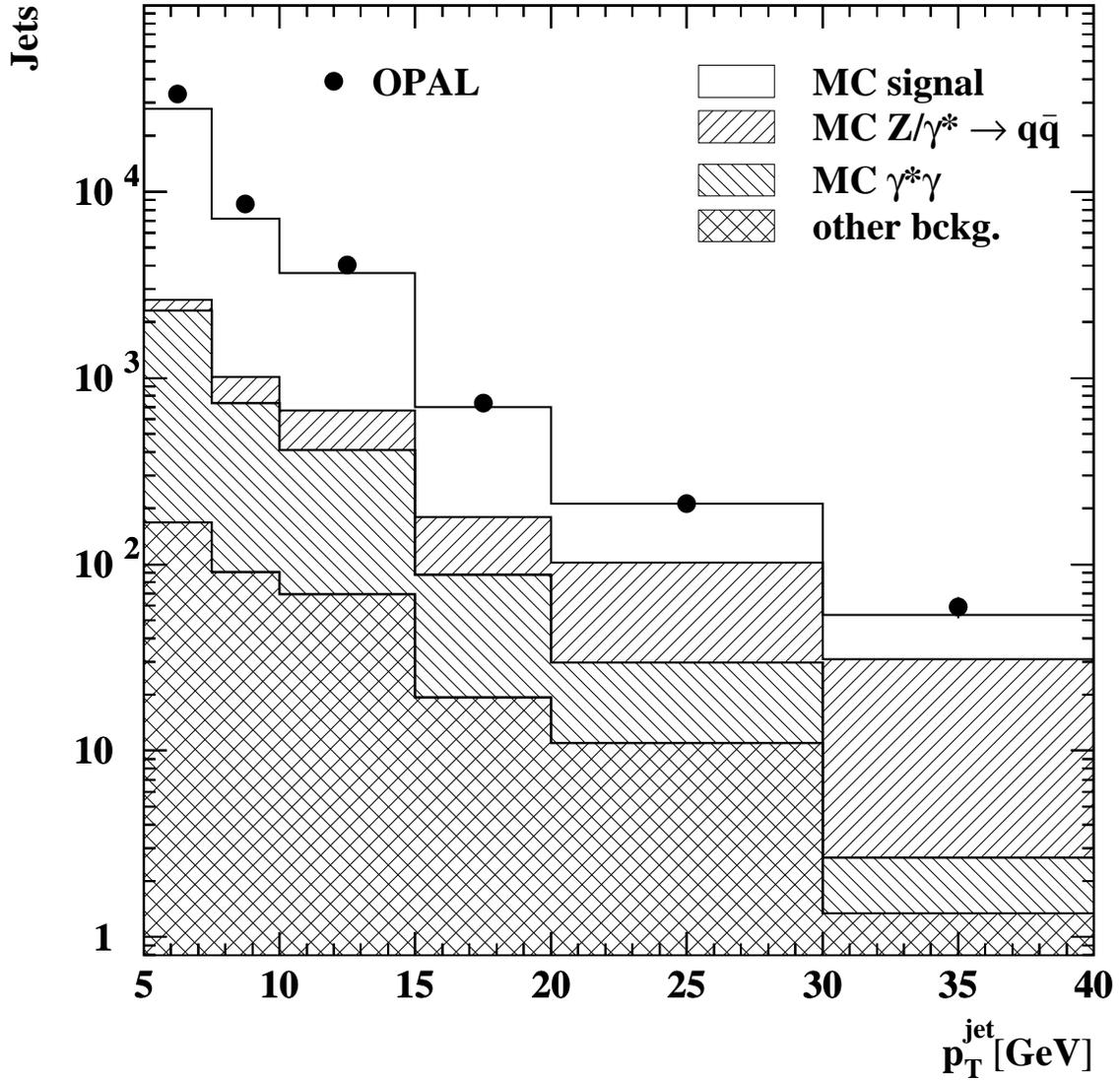}
\end{center}
\caption{Number of jets in each {\ptjet} bin after the event
  selection. Monte Carlo distributions are normalised to the
  luminosity of the data. The statistical uncertainty of the data
  points is shown where larger than the marker size.}
\label{fig:ptbackground}
\end{figure}

The {\ptjet} distribution after the event selection is shown in
Figure~\ref{fig:ptbackground}. The dominant background at low {\ptjet}
is due to {\gsg} events, while for high {\ptjet} the
background is dominated by $\rm{Z}/\gamma^\star \rightarrow
\text{q}\bar{\text{q}}$ events.  To measure the cross-section, the
background is subtracted bin-by-bin.

The measured transverse momentum distributions still have to be
corrected for the losses due to event and track selection cuts, the
acceptance and the resolution of the detector.  This is done using the
PYTHIA~5.722 signal Monte Carlo events which were processed by the
full detector simulation and reconstruction chain.  The data are
corrected by multiplying the experimental distribution with correction
factors which are calculated as the bin-by-bin ratio of the generated
and the reconstructed Monte Carlo distributions.  This method only
yields reliable results if migration effects between bins due to the
finite resolution of the measurement are small. The bins of the
{\ptjet}-distribution have therefore been chosen to be significantly
larger than the detector resolution, obtained from the Monte Carlo
simulation.

\section{Systematic uncertainties}
\label{sec:systematic}

The systematic uncertainties for each bin in $\ptjet$ can be found in
Table~\ref{tab:systematic}.  The values for each bin were averaged
with the results from its two neighbours (single neighbour for
endpoints) to reduce the effect of bin-to-bin fluctuations. The
sources of systematic uncertainties considered are given below.  The
total systematic uncertainty for each bin is calculated by adding the
contributions in quadrature.

\begin{itemize}

\item The absolute energy scale of the ECAL is known to about
  3\%~\cite{bib:totxs} for the jet energy range in this analysis. To
  estimate the influence of this uncertainty, the energy scale of the
  data is varied by this amount, and the analysis is repeated.  

\item To assess the uncertainty associated with the subtraction of
  background events, all backgrounds -- except for {\gsg} -- have been
  varied by $10\%$. The prediction of the contribution
  from {\gsg} events has been scaled up by a factor of two, as
  described earlier. By comparing the MC predictions in regions where
  this background dominates we conclude that this scaling factor can
  not be varied by more than about 30\% while keeping a good
  description of the data. The scaling factor is varied
  accordingly. The resulting uncertainty is dominated by the
  distributions from {\gsg} and $\rm{Z}/\gamma^\star \rightarrow
  \text{q}\bar{\text{q}}$ background.

\item To test the event selection's dependence on the simulation of
  the signal, the signal MC has been re-weighted to have a
  {\ptjet}-slope in which it significantly either over- or
  underestimates the data at high {\ptjet} and the analysis has been
  repeated. The difference between using the original MC and the
  re-weighted MC is included in the systematic uncertainty.

\item The cut on the likelihood output value is varied down to 0.23
  and up to 0.29 for the low {\ptjet} region and down to 0.88 for the
  high $\ptjet$ region.

\item The uncertainty on the determination of the integrated
  luminosity is less than 1\%, and is neglected.

\end{itemize}

\section{Multiple parton interactions and hadronisation corrections}
\label{sec:hcorr}

The NLO calculations do not take into account the possibility of an
underlying event, which leads to an increased energy flow and
therefore to a larger cross-section above any given threshold in the
jet transverse momentum. PYTHIA~6.221 has been used to study the
effect of either considering (default) or leaving out multiple
interactions for the signal MC. In PYTHIA the underlying event is
modelled by multiple parton interactions (MIA). At the lowest
transverse momenta considered the signal MC cross-section increases by
up to 20\% when including MIA.  This effect reduces to less than 10\%
for transverse momenta larger than 7~GeV.

The measured inclusive jet cross-section will be compared to NLO QCD
calculations which describe jet cross-sections for partons, while the
experimental cross-section is presented for hadrons. There is as yet
no rigorous way to use the MC generators to correct the NLO
predictions for this process so that they can be compared to the data,
because the partons in the MC generators and the partons in the NLO
calculations are defined in different ways. But because the use of MC
generators is the only available option so far, they are used to
approximate the size of this hadronisation correction. Hadronisation
corrections have been estimated with PYTHIA 6.221\footnote{The use of
the cluster fragmentation model as implemented in HERWIG as an
alternative in studies carried out in~\cite{bib:dijets} for similar
kinematic conditions yields corrections compatible with or smaller
than those determined using PYTHIA.}. At $\ptjet=$~5~GeV the
correction is about 15\%. The correction decreases with increasing
{\ptjet} and is below 5\% in our study for $\ptjet>$~20~GeV. Disabling
MIA in PYTHIA while determining the hadronisation corrections leads to
values of the correction factors within 2\% of those determined with
MIA enabled.

\section{Differential cross-section}
\label{sec:xsection}

Inclusive jet cross-sections have been measured for the
photon-photon kinematical region of invariant masses of the hadronic
final state $W>5$~GeV, and a photon virtuality
$Q^2 <$~4.5~GeV$^2$. The data are compared to predictions of PYTHIA~6.221
and NLO perturbative QCD~\cite{bib:klasen, bib:frixione}.

The NLO cross-sections are calculated using the QCD partonic
cross-sections in NLO for direct, single- and double-resolved
processes, convoluted with the Weizs\"acker-Williams effective photon
distribution. The hadronisation corrections discussed in the previous
section are applied to the NLO calculation before it is compared to
the data. The GRV-G HO parametrisation of the parton densities of the
photon~\cite{bib:grv} is used with $\Lambda^{(5)}_{\mathrm{DIS}}=
131$~MeV. The renormalisation and factorisation scales in the
calculation are set equal to {\ptjet}.  The cross-section calculations
were repeated for the kinematic conditions of the present
analysis. The calculations shown below are obtained
from~\cite{bib:klasen}.  We have verified that using the independent
calculation presented in~\cite{bib:frixione} yields results within
5\%, except in the lowest bin in {\ptjet}, where it predicts a
cross-section about 25\% higher.

\begin{figure}[ht]
\begin{center}
\includegraphics[width=0.95\textwidth]{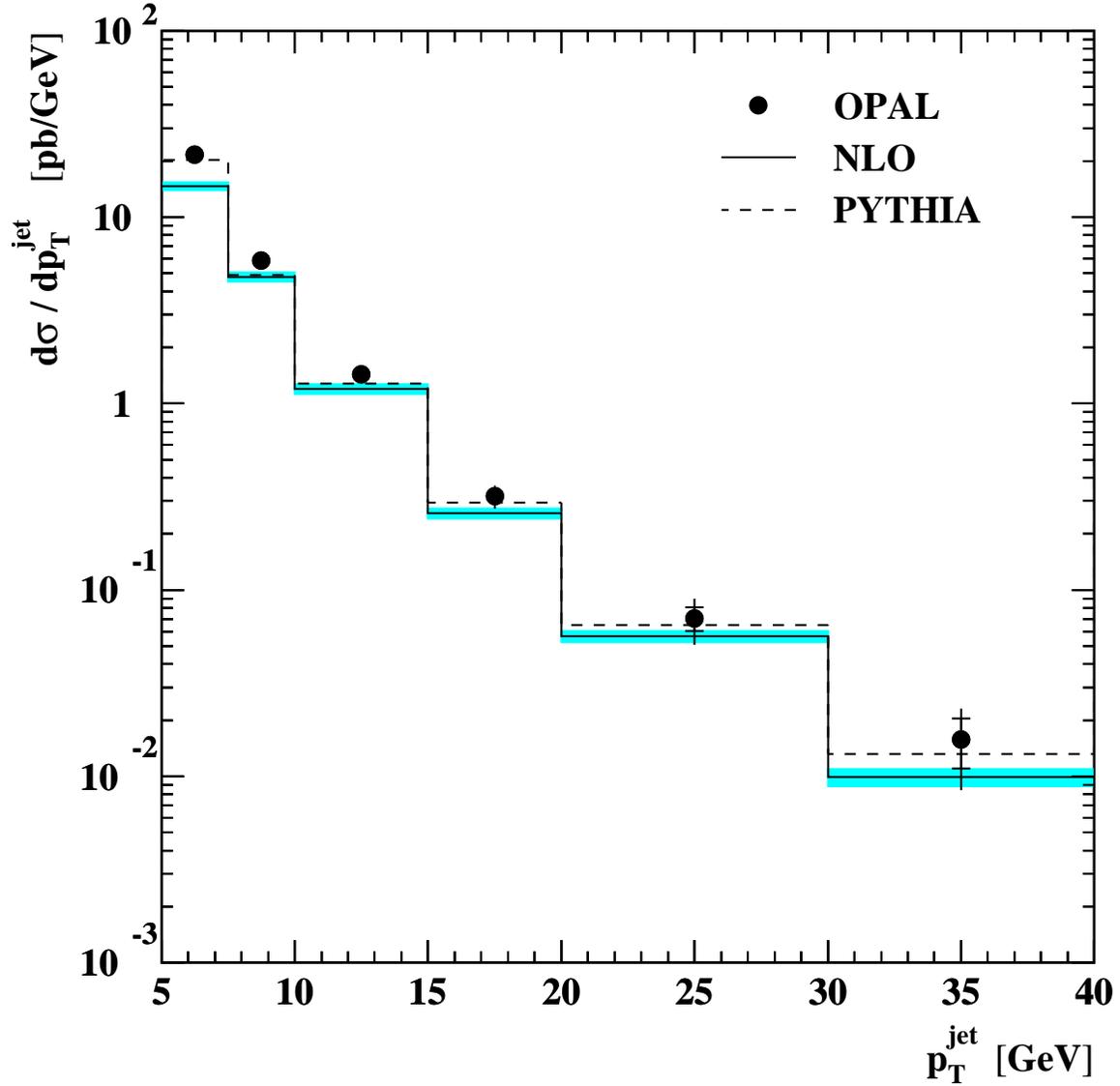}
\end{center}
\caption{Inclusive jet differential cross-section, $d\sigma/d\ptjet$,
  for all jets with $\etajet < 1.5$ compared to NLO and PYTHIA~6.221
  predictions. The hadronisation corrections described in
  section~\ref{sec:hcorr} have been applied to the NLO results. The
  total of statistical and systematic uncertainties added in
  quadrature is shown where larger than the marker size. The inner
  error bars show the statistical uncertainties.  The band on the NLO
  shows the uncertainty associated to the variation of the
  renormalisation and factorisation scale.}
\label{fig:ptcsec}
\end{figure}

Figure~\ref{fig:ptcsec} and Table~\ref{tab:results} show the cross-section
as a function of {\ptjet} for $\etajet < 1.5$. Both PYTHIA~6.221 and the NLO
calculation achieve a good description of the data, with the exception
of the lowest bin in {\ptjet}, where the NLO calculation is too low.

\begin{figure}[ht]
\begin{center}
\includegraphics[width=0.95\textwidth]{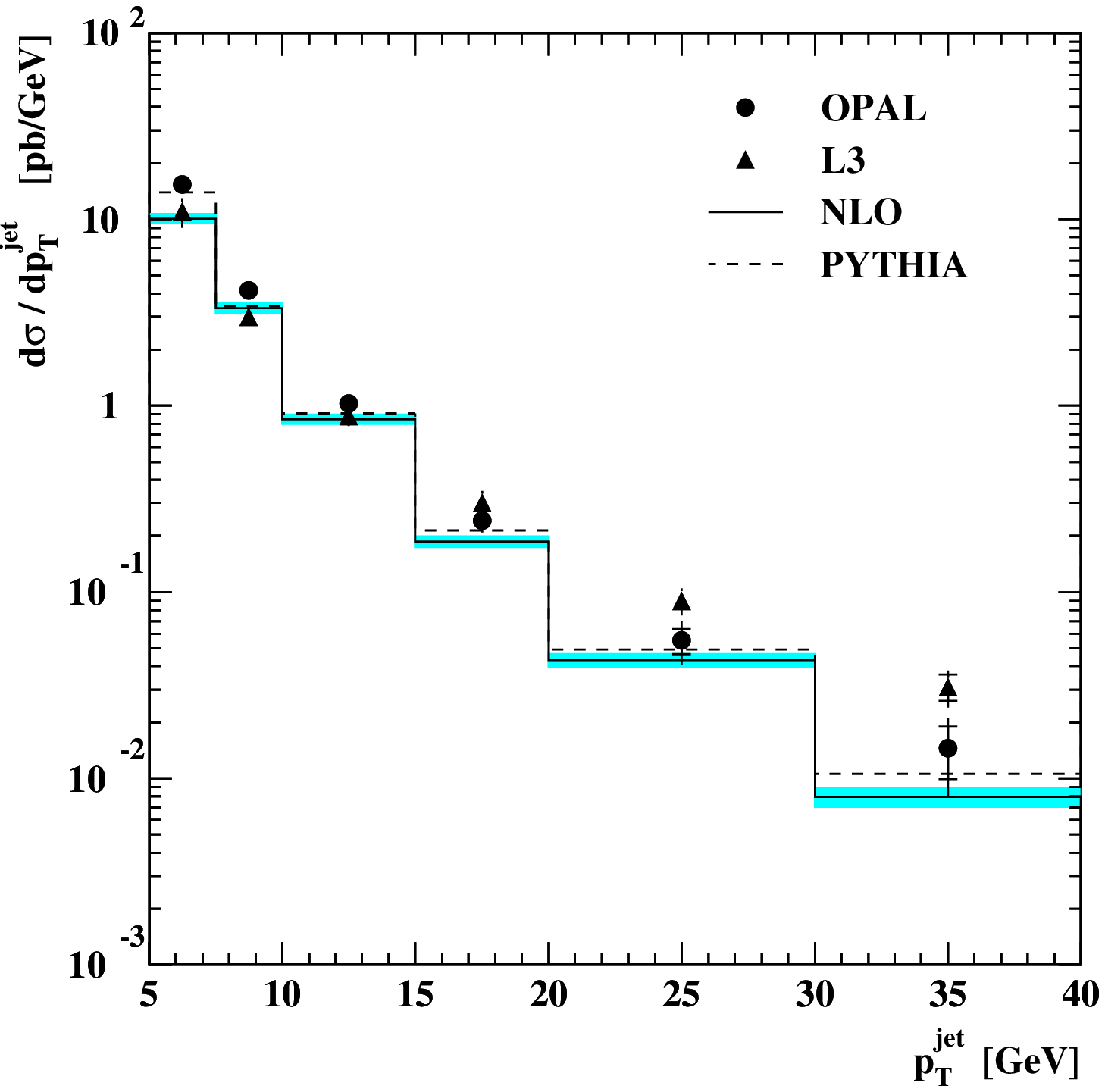}
\end{center}
\caption{Inclusive jet differential cross-section, $d\sigma/d\ptjet$,
  for all jets with $\etajet < 1.0$ compared to the results of the L3
  collaboration, NLO and PYTHIA~6.221 predictions. The hadronisation
  corrections described in section~\ref{sec:hcorr} have been applied
  to the NLO results. The total of statistical and systematic
  uncertainties added in quadrature is shown where larger than the
  marker size. The inner error bars show the statistical
  uncertainties.  The band on the NLO shows the uncertainty associated
  to the variation of the renormalisation and factorisation scale.}
\label{fig:ptcl3}
\end{figure}

To facilitate a comparison with a recent measurement by the L3
collaboration, a measurement of the same quantity as presented in
Figure~\ref{fig:ptcsec} is shown in Figure~\ref{fig:ptcl3} for
$\etajet < 1.0$. While the L3 data points are compatible with the
present measurement, they lie below the OPAL data points at low
{\ptjet} and above the OPAL data points at high {\ptjet}, and there is a
discrepancy in shape between the L3 data and the NLO
calculation. This difference in shape has been reported in the L3
publication and leads to a significant disagreement between the L3 data
and the NLO calculation at the highest {\ptjet} of up to 50~GeV
studied in~\cite{bib:l3paper}. The present analysis finds the
region of ${\ptjet}>40$~GeV to be dominated by background and hence no
measurement is presented for this region.

In contrast to the conclusion in~\cite{bib:l3paper}, the present
analysis finds good agreement between data and calculations for
{\ptjet} of up to 40~GeV, leading to the conclusion that perturbative
QCD in NLO is adequate to describe the process under study.

\section*{Acknowledgements}

We thank Michael Klasen and Stefano Frixione for providing the NLO calculations and for 
many useful discussions.
We particularly wish to thank the SL Division for the efficient operation
of the LEP accelerator at all energies
 and for their close cooperation with
our experimental group.  In addition to the support staff at our own
institutions we are pleased to acknowledge the  \\
Department of Energy, USA, \\
National Science Foundation, USA, \\
Particle Physics and Astronomy Research Council, UK, \\
Natural Sciences and Engineering Research Council, Canada, \\
Israel Science Foundation, administered by the Israel
Academy of Science and Humanities, \\
Benoziyo Center for High Energy Physics,\\
Japanese Ministry of Education, Culture, Sports, Science and
Technology (MEXT) and a grant under the MEXT International
Science Research Program,\\
Japanese Society for the Promotion of Science (JSPS),\\
German Israeli Bi-national Science Foundation (GIF), \\
Bundesministerium f\"ur Bildung und Forschung, Germany, \\
National Research Council of Canada, \\
Hungarian Foundation for Scientific Research, OTKA T-038240, 
and T-042864,\\
The NWO/NATO Fund for Scientific Research, the Netherlands.\\

%
%

\cleardoublepage

\renewcommand{\arraystretch}{1.20}
\begin{table}[t]
\begin{center}
\begin{tabular}{|ccc|c|c|c|c|c|}

\hline

\multicolumn{3}{|c|}{{\ptjet}} & ECAL & Background & Cut &
Signal & Total\\
\multicolumn{3}{|c|}{[GeV]} & energy [\%] & subtraction [\%] &
selection [\%] & rew. [\%] & [\%]\\

\hline

   5.0  &--&  7.5  &    3.2  &   4.4  &  0.1  &  2.6  & 6.0\\
   7.5  &--& 10.0  &    3.5  &   4.6  &  0.2  &  2.2  & 6.2\\
  10.0  &--& 15.0  &    3.6  &   5.3  &  0.8  &  1.4  & 6.6\\
  15.0  &--& 20.0  &    3.7  &   6.2  &  1.7  &  3.1  & 8.0\\
  20.0  &--& 30.0  &    9.1  &   7.7  &  3.7  &  4.0  & 13.2\\
  30.0  &--& 40.0  &   12.2  &   8.6  &  4.7  &  5.0  & 16.5\\

\hline

\end{tabular}
\caption{Systematic uncertainties on the inclusive jet cross-section in the
individual $\ptjet$ bins for $\etajet <$~1.5. Values for $\etajet
<$~1.0 are similar.}
\label{tab:systematic}
\end{center}
\end{table}

\renewcommand{\arraystretch}{1.20}
\begin{table}[t]
\begin{center}
\begin{tabular}{|ccc|l|c|rl|}

\hline

\multicolumn{3}{|c|}{{\ptjet}} & \multicolumn{1}{|c|}{$\avptjet$} & Background &
   \multicolumn{2}{|c|}{$\mathrm{d}\sigma/\mathrm{d}\ptjet$} \\
\multicolumn{3}{|c|}{[GeV]}    & \multicolumn{1}{|c|}{[GeV]}      &   [\%]     &
   \multicolumn{2}{|c|}{[pb/GeV]} \\

\hline
\hline

\multicolumn{7}{|c|}{$\etajet < 1.0$} \\

\hline

   5.0  &  --  &  7.5 & 5.9 &  13.8~$\pm$~0.1  &  (15.3~$\pm$~0.1~$\pm$~0.9) & \\

   7.5  &  --  & 10.0 & 8.5 &  17.4~$\pm$~0.3  &  (41.5~$\pm$~0.8~$\pm$~2.4)  & $\times10^{-1}$ \\

  10.0  &  --  & 15.0 & 11.8 &  21.6~$\pm$~0.4  &  (10.3~$\pm$~0.3~$\pm$~0.6)  & $\times10^{-1}$ \\

  15.0  &  --  & 20.0 & 16.9 &  28.8~$\pm$~0.9  &  (24.1~$\pm$~1.6~$\pm$~1.6)  & $\times10^{-2}$ \\

  20.0  &  --  & 30.0 & 23.3 &  47.6~$\pm$~1.8  &  (55.0~$\pm$~8.4~$\pm$~6.2)  & $\times10^{-3}$ \\

  30.0  &  --  & 40.0 & 33.0 &  57.0~$\pm$~3.6  &  (14.5~$\pm$~4.5~$\pm$~2.0)  & $\times10^{-3}$ \\

\hline
\hline

\multicolumn{7}{|c|}{$\etajet < 1.5$} \\

\hline

   5.0  &  --  &  7.5 & 5.9  &  14.9~$\pm$~0.1  &  (21.7~$\pm$~0.2~$\pm$~1.3) & \\

   7.5  &  --  & 10.0 & 8.5  &  19.3~$\pm$~0.2  &  (58.5~$\pm$~0.9~$\pm$~3.6) & $\times10^{-1}$ \\

  10.0  &  --  & 15.0 & 11.8 &  22.5~$\pm$~0.4  &  (14.3~$\pm$~0.3~$\pm$~0.9) & $\times10^{-1}$ \\

  15.0  &  --  & 20.0 & 16.9 &  28.9~$\pm$~0.9  &  (31.8~$\pm$~1.9~$\pm$~2.6) & $\times10^{-2}$ \\

  20.0  &  --  & 30.0 & 23.5 &  47.1~$\pm$~1.6  &  (70.3~$\pm$~10.2~$\pm$~9.3) & $\times10^{-3}$ \\

  30.0  &  --  & 40.0 & 33.0 &  57.1~$\pm$~3.2  &  (15.7~$\pm$~4.7~$\pm$~2.6) & $\times10^{-3}$ \\

\hline

\end{tabular}
\caption{Background fraction and inclusive jet cross-section for
$\etajet <$~1.0 and $\etajet <$~1.5 as a function of $\ptjet$.  For
the cross-section values the first uncertainty is statistical, the
second is systematic. The uncertainty given for the background
fraction is statistical only.  The average value of $\ptjet$,
$\avptjet$, is also given.}
\label{tab:results}
\end{center}
\end{table}


%
\end{document}